\documentclass[onecolumn,secnumarabic,amsmath,amssymb,balancelastpage,nofootinbib]{article}

\usepackage[e]{esvect}
\usepackage{color}         
\usepackage{graphics}      
\usepackage{graphicx}      
\usepackage{epsf}          
\usepackage{bm}            

\usepackage{natbib}
\usepackage{amssymb}
\usepackage{amsmath, esint}
\usepackage{mathrsfs}
\usepackage{framed}
\usepackage{bigints} 
\usepackage{enumitem}
\usepackage{pifont}
\usepackage[capbesideposition={left,center},facing=yes,capbesidewidth=8cm,capbesidesep=quad]{floatrow} 
\usepackage{setspace} 

\usepackage[none]{hyphenat} 

\usepackage[colorlinks=true]{hyperref}  

\definecolor{darkred}{rgb}{0.6,0,0}
\definecolor{darkgreen}{rgb}{0,0.5,0}
\definecolor{darkblue}{rgb}{0,0,0.6}
\hypersetup{ colorlinks,
linkcolor=darkblue,
filecolor=darkgreen,
urlcolor=darkgreen,
citecolor=darkred }

\newcommand{\del}[0]{\ensuremath{\vec{\nabla}}}

\begin{document}

\sloppy 

\bibliographystyle{authordate1}

\title{\vspace*{-35 pt}\Huge{The Fundamentality of Fields}}
\author{Charles T. Sebens\\Division of the Humanities and Social Sciences\\ California Institute of Technology}
\date{\vspace*{-6 pt}arXiv v.3\ \ \ \ \ September 4, 2022\vspace*{10 pt}\\Forthcoming in \emph{Synthese}}

\maketitle
\vspace*{-20 pt}
\begin{abstract}
There is debate as to whether quantum field theory is, at bottom, a quantum theory of fields or particles.  One can take a field approach to the theory, using wave functionals over field configurations, or a particle approach, using wave functions over particle configurations.  This article argues for a field approach, presenting three advantages over a particle approach: (1) particle wave functions are not available for photons, (2) a classical field model of the electron gives a superior account of both spin and self-interaction as compared to a classical particle model, and (3) the space of field wave functionals appears to be larger than the space of particle wave functions.  The article also describes two important tasks facing proponents of a field approach: (1) legitimize or excise the use of Grassmann numbers for fermionic field values and in wave functional amplitudes, and (2) describe how quantum fields give rise to particle-like behavior.
\end{abstract}

\tableofcontents
\newpage

\section{Introduction}\label{introsec}

What replaces the wave functions of non-relativistic quantum mechanics in relativistic quantum field theory?  According to what I will call the ``particle approach,'' we keep wave functions over particle configurations but modify the formalism to allow for (at least the appearance of) particle creation and annihilation---perhaps by moving from wave functions over configurations of a fixed number of particles to wave functions that assign amplitudes to different numbers of particles being in various possible configurations.  According to the alternative ``field approach,'' we replace wave functions over particle configurations with wave functionals over field configurations---viewing quantum field theory as a theory of fields in quantum superpositions of different classical states.  The debate between these approaches is a debate as to whether fields or particles are more fundamental.\footnote{In \citet{sebensAeon} I give a non-technical introduction this debate.}

The general view seems to be that fields are more fundamental than particles.  But, wave functionals are rarely used\footnote{In his recent book \emph{Foundations of Modern Physics}, \citet[pg.\ 251--252]{weinberg2021} writes: ``Even though it is not generally useful to do so, we can also introduce wave functions for fields -- they are functionals of the field, quantities that depend on the value taken by the field at every point in space, equal to the component of the state vector in a basis labeled by these field values.''} and the field approach (as described above) is rarely explicitly defended.\footnote{\citet{hobson2013}, for example, argues that fields are more fundamental than particles without mentioning wave functionals.}  My goal here is to argue for the field approach, giving a number of reasons to favor the approach that I see as particularly compelling and also noting a few problems for the approach that I see as worthy of attention.  We will be hovering high above some difficult technical terrain, taking a bird's-eye view and pointing elsewhere for elaboration.  Even as a zoomed-out survey, the treatment will be incomplete.  I am only giving a selection of the many considerations that might be offered for and against the field approach.\footnote{Notable omissions include particle localizability, renormalization, spontaneous symmetry breaking, and the Unruh effect (see \citealp{barrett2002, lupher2010, earman2011unruh, ruetsche2011, myrvold2015, baker2016, wallace2021, wallace2022} and references therein).}

In my assessment, the field approach is more attractive than the particle approach.  But, I do not think that the case for the field approach can yet be considered decisive.  There is still work to be done in developing and defending the approach.  I hope to attract allies to those efforts.

Up to now, the particle and field approaches have been debated most carefully in the literature on Bohmian quantum field theory, where one seeks to solve the quantum measurement problem by adding something to the quantum state (perhaps particles, perhaps fields).  However, the debate between particle and field approaches crosscuts the debate as to the right way of solving the quantum measurement problem.  A defender of the many-worlds interpretation might take a unitarily evolving wave functional as fundamental or a unitarily evolving wave function over different numbers and arrangements of particles.  A proponent of collapse could modify the evolution of either kind of quantum state.  To fully understand the ontology of quantum field theory (what exists according to the theory) and the laws of quantum field theory, we need to combine a particle or field approach with a particular solution to the measurement problem.  As a first step, we can set the measurement problem aside and consider the merits of the particle and field approaches.

In general, one might attempt to take a particle or field approach to the entirety of the standard model.  To simplify the discussion here, we will focus on quantum electrodynamics, understood either as a theory of electrons, positrons, and photons, or, alternatively, as a theory of the quantum Dirac and electromagnetic fields.  Although we will primarily be comparing a pure particle approach to a pure field approach, we will also consider a mixed approach, where one treats electrons and positrons as particles interacting with a quantum electromagnetic field, and a deflationary approach, where one views particle wave functions and field wave functionals as equivalent ways of representing the same states.  The pure particle, pure field, mixed, and deflationary approaches are not the only options for understanding states in quantum field theory.  There are other proposals that involve neither wave functions nor wave functionals, but they will not be considered here.  Because some of these other proposals might legitimately claim to be called ``particle'' or ``field'' approaches, one could be more specific and call the two main approaches discussed here ``particle wave function'' and ``field wave functional'' approaches.

This article is organized as follows:  Sections \ref{PARTICLEsection} and \ref{FIELDsection} introduce the particle and field approaches to quantum field theory.  Section \ref{PROSsection} presents three points in favor of the field approach.  First, the particle approach is not available for photons because we do not have a relativistic single-photon quantum theory to build from (like Dirac's single-electron relativistic quantum mechanics).  Second, the classical pre-quantization starting point for the field approach (where the electron is modeled as a rotating cloud of energy and charge in the classical Dirac field) gives a superior account of both spin and self-interaction as compared to the classical pre-quantization starting point for the particle approach (where the electron is modeled as a point particle with intrinsic angular momentum and magnetic moment).  Third, the particle approach appears to have a smaller space of states than the field approach and to lack the states necessary to represent ground and excited states in the presence of interactions.  Section \ref{CONSsection} begins by reviewing some gaps in the field approach that become apparent when articulating the aforementioned advantages and then presents two additional problems facing the field approach.  First, in order to achieve the standard anticommutation relations for field operators we seem forced to use anticommuting Grassmann numbers, both as classical field values and in wave functional amplitudes.  The use of Grassmann numbers leads to problems defining energy and charge densities (in classical field theory) and probability density (in quantum field theory).  Second, there is work to be done in explaining how field wave functionals give rise to particle-like behavior. Section \ref{CONCLUSIONsection} gives a brief conclusion.

\section{The Particle Approach}\label{PARTICLEsection}

\begin{figure}[htb]
\center{\includegraphics[width=9 cm]{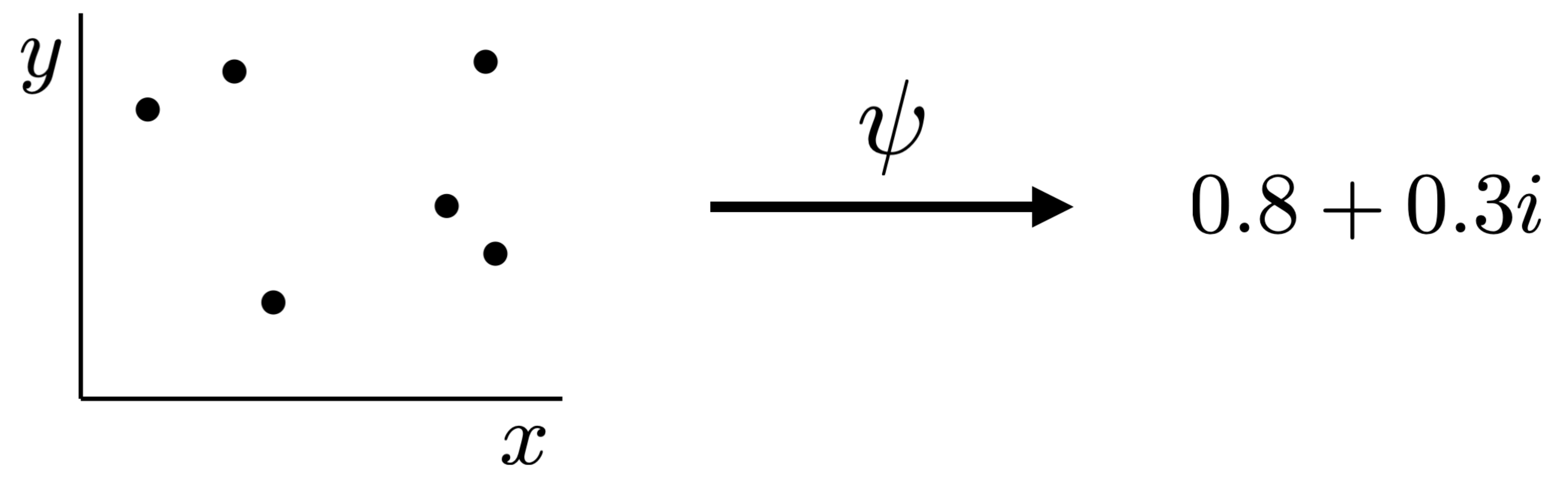}}
\caption{In the particle approach to quantum field theory, the quantum state can be represented as a wave function that takes as input a particular arrangement of some number of particles in space (here depicted as two-dimensional) and returns as output a complex amplitude.}
  \label{particlewf}
\end{figure}

\begin{figure}[htb]
\center{\includegraphics[width=13 cm]{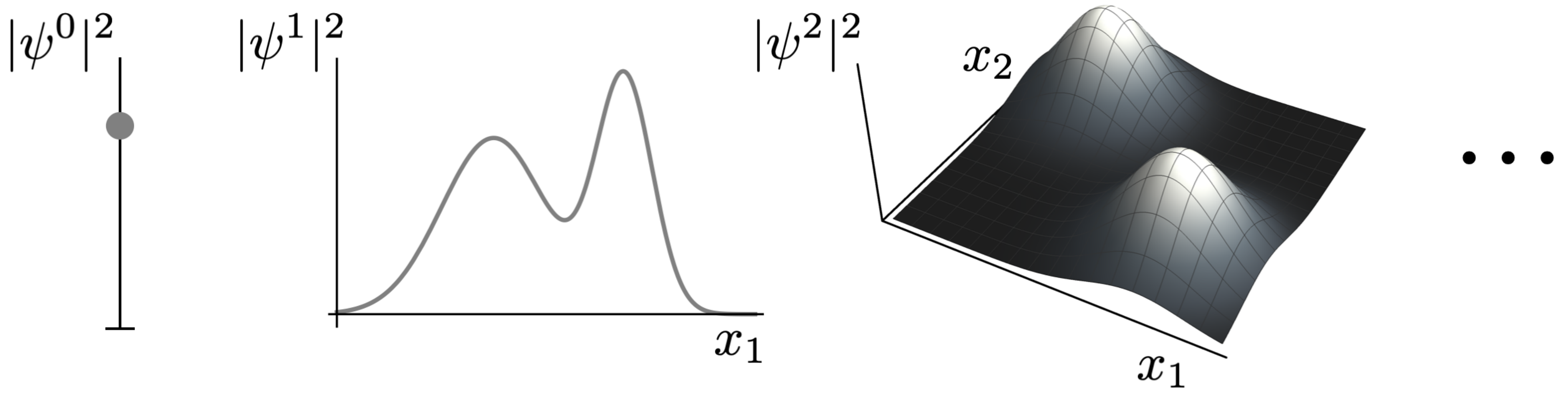}}
\caption{This figure shows the amplitude-squared of a wave function for a variable number of particles, giving a probability density in the collection of $n$-particle configuration spaces for a single spatial dimension.  Here one can see a probability for there being no particles, a probability density for a single particle being in different possible locations, and a probability density for two particles being in different arrangements (that is symmetric under permutation).}
  \label{config}
\end{figure}

For simplicity, let us begin by considering a single boson without spin.  In non-relativistic quantum mechanics, the quantum state of this particle can be given by a wave function assigning complex numbers to points in space that will change over time, $\psi(\vec{x}, t)$ (or, alternatively, by assigning complex numbers to points in momentum space).  For $N$ identical bosons, the quantum state can be given by a symmetric wave function on $N$-particle configuration space: $\psi(\vec{x}_1, \dots, \vec{x}_N, t)$.  As that wave function evolves, particles can interact but the total number of particles will never change.  To find a representation of the quantum state better suited to relativistic quantum field theory (where we have particle creation and annihilation), we can introduce a wave function that spans the various different $n$-particle configuration spaces,
\begin{equation}
\psi \left\{ \begin{matrix}
\psi^{0}(t) \\
\psi^{1}(\vec{x}_1,t) \\
\psi^{2}(\vec{x}_1,\vec{x}_2,t) \\
\vdots
\end{matrix}
\right.\ .
\label{totalWFscalar}
\end{equation}
The total wave function is composed of a $0$-particle wave function, a $1$-particle wave function, a (symmetric) $2$-particle wave function, and so on.\footnote{This kind of particle approach is described in \citet[sec.\ 6f, 6h, and 7c]{schweberQFT}; \citet{dgz2004, durr2005, tumulka2018}.  Although the focus here is on wave functions that assign amplitudes to different particle arrangements at a single time, some have proposed (for better harmony with special relativity) using multi-time wave functions where there is a separate time coordinate for each particle position (\citealp{lienert2017}; \citealp[ch.\ 4]{lienert2020}).}  The total wave function assigns complex amplitudes to all points in the disjoint union of $n$-particle configuration spaces (figure \ref{particlewf}).  The amplitude-squared of the wave function gives a probability density in this space (figure \ref{config}).  In terms of the appropriate particle creation operators $a^\dagger (\vec{x})$, this quantum state can be written as
\begin{equation}
| \psi (t) \rangle = \left(\psi^{0}(t) + \int d^3 x_1 \psi^{1}(\vec{x}_1,t) a^\dagger (\vec{x}_1) + \frac{1}{\sqrt{2}}\int d^3 x_1 d^3 x_2 \psi^{2}(\vec{x}_1,\vec{x}_2,t) a^\dagger (\vec{x}_2) a^\dagger (\vec{x}_1)+\dots \right)| 0 \rangle
\label{WFscalar}
\end{equation}
Alternatively, one can Fourier transform and express the state in terms of creation operators for particular momenta.\footnote{Some authors present the particle approach for momenta as a potentially viable option (at least in the absence of interactions), but challenge the idea that Fourier transforming yields a relativistically acceptable representation in terms of positions.  See \citet[pg.\ 48--56, 85--91]{teller}; \citet{myrvold2015}.}  The space of all possible quantum states \eqref{WFscalar} for a variable number of particles is a ``Fock space.'' The dynamics for the quantum state \eqref{WFscalar} can be given by a Schr\"{o}dinger equation of the general form,
\begin{equation}
i \hbar \frac{d}{dt} | \psi (t) \rangle = \widehat{H} | \psi (t) \rangle
\ .
\label{generalschrodinger}
\end{equation}
Depending on the Hamiltonian $\widehat{H}$, we can incorporate our wave function for a variable number of particles \eqref{totalWFscalar} into either a relativistic or a non-relativistic quantum field theory.  Our focus here will be on relativistic quantum field theory.

For electrons and positrons, things are more complicated.  There are two broad strategies for developing a particle approach: one involving a variable number of particles and the other invoking the Dirac sea.  The first option is described in \citet[sec.\ 8b]{schweberQFT}; \citet[sec.\ 10.1]{thaller1992}; \citet[sec.\ 11.3.1]{durr2020}.  We can begin with a four-component complex-valued wave function $\psi_i(\vec{x},t)$ obeying the free Dirac equation (where $i$ is an index on the four components).  Just as a single-component wave function obeying the Schr\"{o}dinger equation in non-relativistic quantum mechanics can be written as a superposition of plane wave momentum eigenstates, a general four-component wave function obeying the Dirac equation can be written as a superposition of plane wave momentum eigenstates (or ``modes'').\footnote{See \citet[ch.\ 3]{bjorkendrell}.}  For a given momentum $\vec{p}$, there are multiple plane wave eigenstates---differing in spin and also in whether the eigenstate is positive frequency, with time dependence $e^{-\frac{i}{\hbar}\mathcal{E}(\vec{p}\,) t}$, or negative frequency, with time dependence $e^{\frac{i}{\hbar}\mathcal{E}(\vec{p}\,) t}$ (where $\mathcal{E}(\vec{p}\,)=\sqrt{m^2c^4+|\vec{p}\,|^2c^2}\,$).  We can interpret a normalized solution of the free Dirac equation $\psi^{(1,0)}_{i_1}(\vec{x}_1,t)$ composed entirely of positive-frequency modes as a single-electron wave function (reserving negative-frequency modes for the representation of positrons).  Because electrons are fermions, multi-electron wave functions, $\psi^{(n,0)}_{i_1 \dots i_n}(\vec{x}_1, \dots, \vec{x}_n,t)$, must be antisymmetric.  We can construct such wave functions by superposing antisymmetric products of positive-frequency modes.  One can also introduce single-positron states, $\psi^{(0,1)}_{j_1}(\vec{y}_1,t)$, composed of negative-frequency modes.  The total wave function can be expressed on the disjoint union of $n$-electron and $m$-positron configuration spaces, as in \eqref{totalWFscalar},
\begin{equation}
\psi \left\{ \begin{matrix}
\psi^{0}(t) \\[5 pt]
\psi^{(1,0)}_{i_1}(\vec{x}_1,t) \\[5 pt]
\psi^{(0,1)}_{j_1}(\vec{y}_1,t) \\[5 pt]
\psi^{(1,1)}_{i_1; j_1}(\vec{x}_1; \vec{y}_1,t) \\[5 pt]
\psi^{(2,0)}_{i_1,i_2}(\vec{x}_1,\vec{x}_2,t) \\[5 pt]
\vdots
\end{matrix}
\right.\ ,
\label{totalWFDirac}
\end{equation}
where each piece $\psi^{(n,m)}_{i_1,\dots,i_n; j_1,\dots,j_m}(\vec{x}_1,\dots,\vec{x}_n; \vec{y}_1,\dots,\vec{y}_m,t)$ is separately antisymmetric under exchange of electrons or positrons.  So far, we have focused on the free Dirac equation.  But, one might hope that this kind of representation can be used when we introduce interactions that might cause the probability density to shift from certain $n$-electron and $m$-positron configuration spaces to others, as particles are created and destroyed.

The second strategy for developing a particle approach is described in \citet[pg.\ 276]{bohmhiley}; \citet{colin2007}; \citet{deckert2020}; \citet[sec.\ 11.3.2]{durr2020}.  This strategy starts again from solutions $\psi_i(\vec{x},t)$ to the free Dirac equation, but allows electrons to enter states composed of both positive-frequency and negative-frequency modes.  The negative-frequency modes are understood to be negative-energy modes that are ordinarily filled (so that, by Pauli exclusion, the states are rendered unavailable).  In standard unbounded three-dimensional space, there would be infinitely many negative-energy modes that would have to be filled by infinitely many electrons (an infinite ``Dirac sea'').  However, if we assume that the volume of space is finite and impose a cutoff on high momenta modes\footnote{These are controversial assumptions.  For more on why such assumptions might be made and on the consequences that follow from them, see the references in footnote \ref{cutofffootnote}.} (as in \citealp{colin2007, deckert2020}), then there are only finitely many negative-energy modes and we can take the true number of electrons to be some fixed number $N$ such that the wave function is simply the antisymmetric
\begin{equation}
\psi^{N}_{i_1,\dots,i_N}(\vec{x}_1,\dots,\vec{x}_N,t)
\ .
\end{equation}
In the ground state, the negative-energy modes are filled by a vast number of electrons.  In excited states, there are electrons in positive-energy modes and unfilled negative-energy modes (holes in the Dirac sea) that act like positively charged particles (positrons).

For photons, the particle approach has trouble getting off the ground because we do not have a relativistic wave equation like the Dirac equation to serve as our starting point.  We will discuss this problem and possible responses in section \ref{PHOTONSsection}.

\section{The Field Approach}\label{FIELDsection}

\begin{figure}[htb]
\center{\includegraphics[width=9 cm]{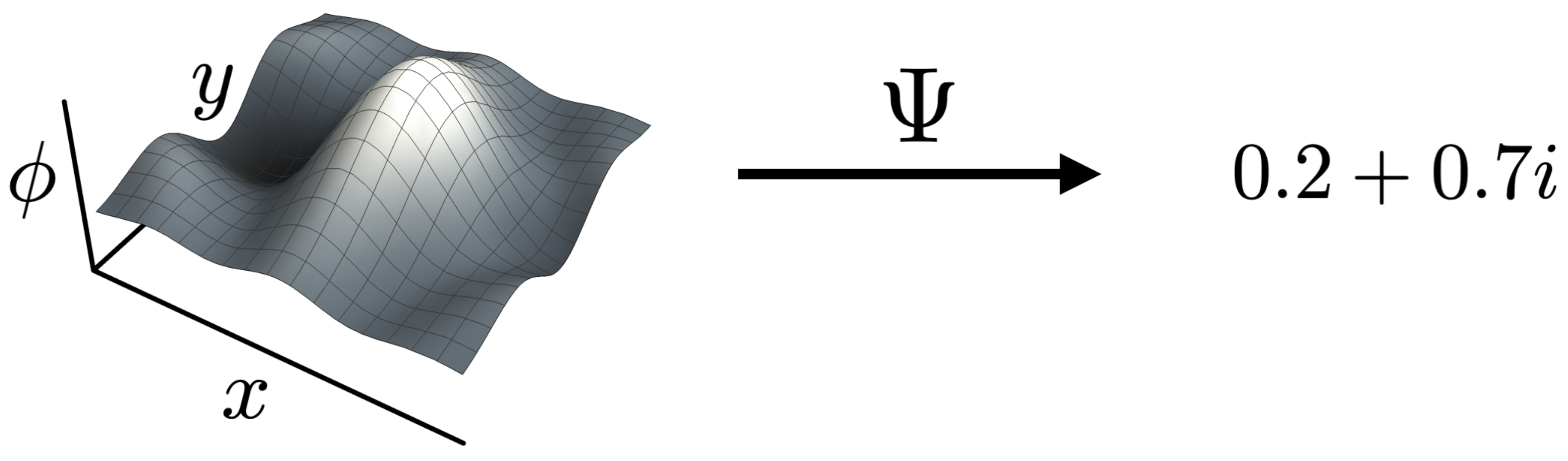}}
\caption{In the field approach to quantum field theory, the quantum state can be represented as a wave functional that takes as input a classical field configuration and returns as output an amplitude.  In this figure, the sole classical field is taken to be a real scalar field $\phi$ in two-dimensional space.}
  \label{fieldwf}
\end{figure}

According to the field approach,\footnote{Detailed technical introductions to the field approach are given in \citet{jackiw1987, jackiw1990, floreanini1988, hatfield}; \citet[ch.\ 11]{bohmhiley}; \citet[sec.\ 12.4]{holland}; \citet{kiefer1994, kaloyerou1994, kaloyerou1996}; \citet[pg.\ 29--33]{huang2008}.  The field approach is also discussed in \citet{valentini1992, valentini1996, huggett2000, wallace2001, wallace2006, wallace2021, baker2009, baker2016, struyve2010, struyve2011, myrvold2015}; \citet[sec.\ 11.2]{durr2020}.  For an introduction to the field approach aimed at a general audience, see \citet[ch.\ 12]{carroll2019}.} quantum field theory should be viewed as a true theory of fields.  Instead of wave functions that assign quantum amplitudes to possible arrangements of point particles, we should use wave functionals that assign quantum amplitudes to possible configurations of classical fields.  The wave functional for a single field takes as input a full classical state of that field at a given moment, specifying its values at every point in space (a classical field configuration).  Because the classical field configuration is itself a function from points in space to field values, the quantum state is a function of a function---called a ``functional.''  In a quantum field theory for a real scalar field $\phi$, the complex-valued wave functional can be written as $\Psi[\phi,t]$ (figure \ref{fieldwf}).  The wave functional's amplitude-squared gives a probability density on the space of all possible field configurations.\footnote{There are mathematical issues regarding the definition of a measure over the (infinite-dimensional) space of possible field configuration---a measure that is necessary for a mathematically rigorous account as to how the amplitude-squared of the wave functional serves as a probability density (see \citealp[sec.\ 2.2.2]{struyve2010}).\label{measurefootnote}}  The dynamics of the wave functional are given by a Schr\"{o}dinger equation,
\begin{equation}
i \hbar \frac{d}{dt} \Psi[\phi,t] = \widehat{H} \Psi[\phi,t]
\ .
\label{superschrodinger}
\end{equation}
By having the state evolve and not the operators, we are working in the Schr\"{o}dinger picture (\citealp{struyve2010, struyve2011} calls this the ``functional Schr\"{o}dinger picture'').  Of course, the use of wave functionals is entirely compatible with the Heisenberg or interaction pictures, should those be preferable for certain purposes (that choice will just alter whether and how the wave functional evolves).  From \eqref{superschrodinger}, one can generate path integral equations for state evolution and derive Feynman rules for perturbatively approximating the evolution in scattering contexts (though we will not do so here).

\begin{figure}[htb]
\center{\includegraphics[width=12 cm]{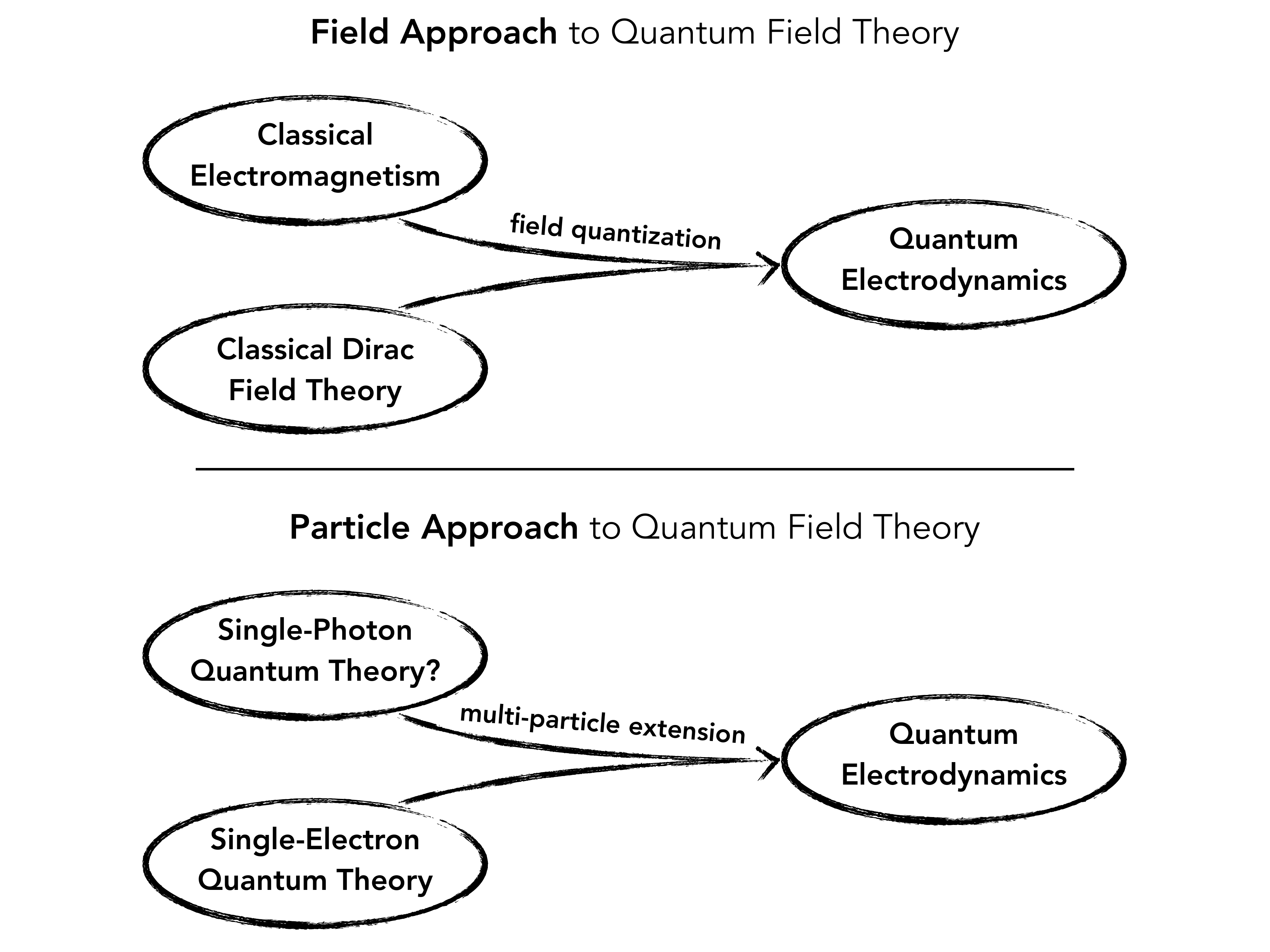}}
\caption{On the field approach, quantum electrodynamics can be arrived at by starting with separate classical theories of the free Dirac and electromagnetic fields, combining these into a single classical theory of interacting Dirac and electromagnetic fields, and then quantizing this classical field theory.  On the particle approach, quantum electrodynamics is viewed as an extension of Dirac's single-electron relativistic quantum mechanics to multiple electrons (and positrons) with something added to capture electromagnetic interactions: either a similar quantum treatment of photons or direct interactions between electrons and positrons.  The field and particle approaches disagree on the route to quantum field theory and on the architecture of the destination---disagreeing as to whether field wave functionals or particle wave functions should be used to represent quantum states.}
  \label{paths}
\end{figure}

Taking a field approach to quantum electrodynamics, our starting point is a classical relativistic theory of interacting Dirac and electromagnetic fields (figure \ref{paths}).\footnote{This classical theory is discussed in \citet{barut1964}; \citet[sec.\ 15.2]{bjorkendrellfields}; \citet[sec.\ 20.9]{doughty1990}; \citet[sec.\ 5.1]{greiner1996}; \citet[sec.\ 8.1]{hatfield}; \citet{potentialenergy}.}  We arrive at quantum electrodynamics by quantizing these fields---allowing them to enter superpositions of classical states described by a wave functional.  In the classical field theory that precedes quantum electrodynamics, the electromagnetic field evolves by Maxwell's equations (with the charge and current densities of the Dirac field acting as source terms) and the Dirac field evolves by the Dirac equation (with the electromagnetic field playing a part in its evolution).  Although the Dirac equation is familiar, it is usually presented as part of a quantum theory (as in section \ref{PARTICLEsection}).  Here, we are viewing the Dirac equation as part of a classical field theory that yields quantum electrodynamics upon field quantization.  The thing that evolves by the Dirac equation in this classical field theory, $\psi_i(\vec{x},t)$, may look like a quantum wave function, but in this context it is to be interpreted as a four-component classical field (the same kind of thing as the classical electromagnetic field).

In a field approach to quantum electrodynamics, we would like a wave functional defined over classical configurations of both the Dirac field (for electrons and positrons) and the electromagnetic field (for photons).  Let us first consider the electromagnetic field on its own.  The usual way of introducing a wave functional\footnote{See \citet[appendix A]{bohm1952pt2}; \citet[sec.\ 10.2]{hatfield}; \citet{kaloyerou1994, kaloyerou1996}; \citet[sec.\ 4.2]{struyve2010}; \citet{flack2016}.} is to adopt the Coulomb gauge or the temporal gauge, on either of which the scalar potential $\phi$ vanishes (if there is no charged matter) and the state of the classical electromagnetic field can be given by the vector potential $\vec{A}$ alone.  The wave functional $\Psi[\vec{A},t]$ assigns complex amplitudes to possible configurations of the vector potential.  The classical energy of the electromagnetic field can be converted into a Hamiltonian operator that gives the evolution of this wave functional via a Schr\"{o}dinger equation like \eqref{superschrodinger}.  The wave functional's amplitude-squared gives a probability density over possible classical configurations of the electromagnetic field.

Next, let us consider the Dirac field on its own.  Wave functionals have also been used for the Dirac field,\footnote{See \citet{floreanini1988, jackiw1990, valentini1992, valentini1996, hatfield, kiefer1994}.} but their application in that context is less elegant.  The problem is that the field operator $\widehat{\psi}_i(\vec{x})$ is ordinarily understood to multiply the wave functional by the value of the $i$-th component of the Dirac field at $\vec{x}$,
\begin{equation}
\widehat{\psi}_i(\vec{x})\Psi[\psi,t]=\psi_i(\vec{x})\Psi[\psi,t]
\ ,
\label{fieldoperator}
\end{equation}
and also ordinarily taken to obey the anticommutation relation
\begin{equation}
\left\{\widehat{\psi}_i(\vec{x},t),\widehat{\psi}_j(\vec{y},t)\right\}=0
\ .
\label{diracanticommutation}
\end{equation}
When we place that anticommutator in front of the wave functional, it gives
\begin{equation}
\left\{\widehat{\psi}_i(\vec{x}),\widehat{\psi}_j(\vec{y})\right\}\Psi[\psi,t]=\left(\psi_i(\vec{x})\psi_j(\vec{y})+\psi_j(\vec{y})\psi_i(\vec{x})\right)\Psi[\psi,t]
\ ,
\end{equation}
which will only be zero if the field values themselves anticommute.  Thus, when wave functionals are used for the Dirac field (or any other fermion field), the field values are taken to be anticommuting Grassmann numbers.  The use of Grassmann numbers is standard for path integrals in quantum field theory,\footnote{See \citet{hatfield}; \citet[sec.\ 9.5]{peskinschroeder}; \citet[sec.\ 6.7]{ryder}; \citet[sec.\ 12.8]{greiner1996}; \citet[sec.\ 11.5]{zee2010}; \citet[sec.\ 10.3.2]{duncan}; \citet[sec.\ 14.6]{schwartz}.} but there are a number of reasons why one might be concerned about their use for Dirac field wave functionals.  We will discuss these concerns in section \ref{GRASSMANNsection}.

\section{Reasons to Favor the Field Approach}\label{PROSsection}

This section presents some advantages of the field approach over the particle approach: we do not have a relativistic single-photon quantum theory from which to develop a particle approach to quantum electrodynamics, a classical field model of the electron is superior to a classical particle model as regards both spin and self-interaction, and the space of states in the particle approach appears to be too small.

\subsection{No Particle Approach for Photons}\label{PHOTONSsection}

For a single electron, we have a relativistic quantum theory where the electron's wave function $\psi_i(\vec{x},t)$ evolves by the Dirac equation,
\begin{equation}
i\hbar \frac{\partial \psi_i}{\partial t}=\left(c\:\vec{\alpha}_{ij}\cdot\vec{p}+\beta_{ij} m c^2 \right)\psi_j
\ ,
\label{thediracequation}
\end{equation}
written here without any interactions.  In \eqref{thediracequation}, the alpha and beta matrices are standard and $\vec{p}=-i\hbar\del$.  The probability and probability current densities are given by\footnote{The indices on $\psi$ are dropped in \eqref{diracdensities} and some other equations.  One could write $\psi^\dagger \psi$ as $\sum_{i=1}^4 \psi_i^*\psi_i$.}
\begin{align}
\rho^{\,p}&=\psi^\dagger \psi
\nonumber
\\
\vec{J}^{\:p}&=c \psi^\dagger \vec{\alpha} \psi
\ .
\label{diracdensities}
\end{align}
This single-electron quantum theory can then be extended to a variable number of electrons and positrons, as outlined in section \ref{PARTICLEsection}.  For the photon, we do not have a similar relativistic single-particle quantum theory with a relativistic wave equation and well-behaved densities of probability and probability current (as would be required for the kind of particle approach described in section \ref{PARTICLEsection}).  There are ways you might try to develop such a theory, but none have been widely regarded as successful.  For our purposes here, it will be worthwhile to review a couple attempts (discussed in \citealp{emasqp}).\footnote{The difficulties involved in constructing a relativistic quantum theory for the photon are reviewed in \citet[sec.\ II.5.2]{bohm1987}; \citet{holland1993}; \citet[sec.\ 12.6]{holland}; \citet{kiessling2018}; \citet[sec.\ 7.4]{valentini2020}.}

Around 1930, both Ettore Majorana and Georg Rumer\footnote{See \citet{rumer1930, mignani1974}.} considered taking the electric field plus $i$ times the magnetic field to be a photon wave function,
\begin{equation}
F_i=E_i+iB_i
\ .
\label{webervector}
\end{equation}
The problem with this wave function is that $F^{\dagger}F$ gives an energy density, not a probability density.  One way to remedy this problem is to follow an idea from \citet{good1957}:\footnote{Similar proposals appear in \citet[eq.\ 1.6]{ab1965}; \citet[pg.\ 191]{pauli}; \citet[pg.\ 637]{mandelwolf}.} Fourier transform the putative wave function in \eqref{webervector}, divide by the square root of the photon energy $\hbar k c$ (where $k$ is the wave number and $\hbar k$ is the momentum), and then Fourier transform back,
\begin{equation}
\phi_i(\vec{x})=\frac{1}{\sqrt{8\pi}}\frac{1}{(2\pi)^{3}}\int{d^3 k\left[\frac{e^{i \vec{k}\cdot\vec{x}}}{\sqrt{\hbar k c}}  \int{d^3 y\: e^{-i \vec{k}\cdot\vec{y}} F_i(\vec{y})}\right] }
\ .
\label{photonWF}
\end{equation}
The candidate photon wave function $\phi_i(\vec{x})$ obeys the wave equation,
\begin{equation}
i\hbar\frac{\partial \phi_i}{\partial t}  = c \: \vec{s}_{ij}\cdot \vec{p}\: \phi_j
\ ,
\label{thephotonequation}
\end{equation}
which can be derived from Maxwell's equations and closely resembles the Dirac equation \eqref{thediracequation}, though (as one would expect) there is no mass term.  The probability and probability current densities for this wave function are given by
\begin{align}
\rho^{\,p}&=\phi^\dagger \phi
\nonumber
\\
\vec{J}^{\:p}&=c \phi^\dagger \vec{s} \phi
\ ,
\label{photonprob}
\end{align}
resembling \eqref{diracdensities}.  In the above equations, the $\vec{s}$ matrices can be expressed in terms of the Levi-Civita symbol as $(s_i)_{jk}=-i \epsilon_{ijk}$.  Unfortunately, the densities in \eqref{photonprob} do not transform properly under Lorentz transformations and this ultimately renders Good's promising idea unacceptable \citep{emasqp}.  The Majorana-Rumer and Good photon wave functions are just two natural proposals.  \citet[appendix A]{kiessling2018} discuss problems for a few other ways one might attempt to introduce a photon wave function.  As things stand, we do not have a widely accepted relativistic quantum mechanics for the photon.  Without such a theory, it is hard to see how we might develop a particle approach for photons in quantum electrodynamics.

In my assessment, the current inability of the particle approach to incorporate photons is a strong mark against it, pointing to the field approach as the more promising direction for understanding states in quantum electrodynamics.  However, there are multiple ways that a proponent of particles might respond:  First, you could see the current situation as a challenge and work to find an acceptable single-photon relativistic quantum theory.\footnote{Recent proposals for such a theory have been given in \citet{kiessling2018, bb2019, hawton2019, hawton2021}.}  Such efforts are worthwhile, and may allow the particle approach to be extended to photons.  However, the failure of certain natural proposals gives us reason to question whether such a theory is there to be found.  Second, you can lower your ambitions and accept a single-photon relativistic quantum theory that does not give densities of probability and probability current in space (though this would not yield the kind of particle approach described in section \ref{PARTICLEsection}).\footnote{\citet[pg.\ 14]{berestetskii1982} write that ``the coordinate wave function of the photon cannot be interpreted as the probability amplitude of its spatial localization.''  (See also \citealp[sec.\ 2.2]{ab1965}.)}  Third, you could adopt a mixed approach to quantum field theory where one takes a particle approach for fermions (including electrons and positrons) and a field approach for bosons (including photons).\footnote{Such mixed approaches are considered in \citet{bohm1987, bohmhiley}; \citet[pg.\ 293]{kaloyerou1994}; \citet[pg.\ 155]{kaloyerou1996}.}  This idea fits well with the common presentations of classical electrodynamics as a theory of charged particles interacting with the electromagnetic field, and would sidestep the challenges facing the field approach when it is applied to fermions (section \ref{GRASSMANNsection}).  Fourth, you might seek to eliminate photons entirely and have charged particles interact directly with one another---understanding both classical and quantum electrodynamics as theories involving action-at-a-distance.\footnote{\citet{lazarovici2018} advocates this kind of approach.}

The classical and quantum equations describing the electromagnetic and Dirac fields are sufficiently similar that I think it is unappealing to adopt different approaches for each.\footnote{\citet[pg.\ 155]{kaloyerou1996} gives a different argument for consistency in the approaches used for bosons and fermions (in the context of seeking a Bohmian quantum field theory):
\begin{quote}
``A criterion that has been introduced by Bohm, regarded as preliminary by the present author, is that where the classical limit of the equation of motion of the field is a wave equation, then the entity can be consistently regarded as an objectively existing field, but where the classical limit is a particle equation, then the entity must be regarded as an objectively existing particle. The former is the case for bosons, such as the electromagnetic field and the mesons, and the latter for fermions. The problem with this criteria is that the field ontology of bosons is in direct conflict with that of fermions when it is recalled that some bosons are fermion composites (e.g., mesons are quark-antiquark pairs) and quarks are fermions. It seems likely instead that fermions and bosons should have the same ontology.''
\end{quote}}  This speaks against the second and third responses described above, where one takes a particle approach for electrons and positrons while taking either a field approach for photons or eliminating them entirely.  There are many ways to see this similarity.  Let us take an unusual approach here and consider \eqref{photonWF} as an alternative way of representing the classical electromagnetic field (instead of viewing it as a candidate photon wave function).  In this notation, the free dynamics for the electromagnetic field \eqref{thephotonequation} closely resemble the free dynamics for the Dirac field \eqref{thediracequation}.  If we decompose the electromagnetic field $\phi$ into a positive-frequency part $\phi_+$ and a negative-frequency part $\phi_-$, the energy of the electromagnetic field can be written (in Gaussian cgs units) as,\footnote{See \citet{good1957}; \citet[sec.\ 6]{positrons}.}
\begin{align}
\mathcal{E}&=\int d^3x\left(\frac{E^2}{8\pi}+\frac{B^2}{8\pi}\right)
\nonumber
\\
&=i\hbar \int d^3x\left(\phi_+^{\dagger}\frac{\partial \phi_+}{\partial t}-\phi_-^{\dagger}\frac{\partial \phi^{\dagger}_-}{\partial t}\right)
\ .
\label{EMenergy}
\end{align}
This closely resembles the standard energy of the free Dirac field,\footnote{This total energy can be calculated from the $00$ component of either the canonical or the symmetrized energy-momentum tensor for the free Dirac field.  See \citet[pg.\ 419]{heitler}; \citet[pg.\ 219]{schweberQFT}; \citet[eq.\ 3]{positrons}.}
\begin{equation}
\mathcal{E}=i\hbar \int d^3x\left(\psi_+^{\dagger}\frac{\partial \psi_+}{\partial t}+\psi_-^{\dagger}\frac{\partial \psi_-}{\partial t}\right)
\ .
\end{equation}
The resemblance can be made even closer if we flip the sign of the energy for the negative-frequency modes,
\begin{equation}
\mathcal{E}=i\hbar \int d^3x\left(\psi_+^{\dagger}\frac{\partial \psi_+}{\partial t}-\psi_-^{\dagger}\frac{\partial \psi_-}{\partial t}\right)
\ .
\end{equation}
This modification has been advocated in \citet{positrons} as a way of altering classical Dirac field theory so that negative-frequency modes represent positive-energy positrons, not negative-energy electrons.  The charge and current densities for the classical Dirac field can also be modified so that negative-frequency modes carry positive charge.  There remains work to be done incorporating these modifications into a theory of interacting Dirac and electromagnetic fields.

\subsection{Comparing Classical Theories: Spin and Self-Interaction}\label{CLASSICALsection}

In either the particle or the field approach to quantum field theory for electrons and positrons, we can view the starting point as the Dirac equation \eqref{thediracequation}.  On the particle approach, this is interpreted as a relativistic quantum equation that gives the dynamics for a single-electron wave function.  Quantum field theory is seen as a multi-particle extension of this single-particle theory.  The wave functions of quantum field theory describe superpositions of classical states where electrons are point particles with definite locations, intrinsic angular momenta, and (oppositely oriented) intrinsic magnetic moments.  On this approach, the classical theory that one would quantize to arrive at quantum field theory is a classical theory of point-size electrons and positrons (or just electrons in a Dirac sea version of the particle approach).

On the field approach, the Dirac equation is interpreted as part of a classical relativistic field theory where it gives the dynamics for the Dirac field.  One then arrives at a quantum field theory for electrons and positrons by quantizing this classical field theory (by field quantization).  The wave functional describes superpositions of classical states where the Dirac field has definite values everywhere (and thus definite densities of charge, current, energy, and momentum).  For now, let us take the classical Dirac field to be complex-valued and leave the possibility of a Grassmann-valued classical Dirac field to section \ref{GRASSMANNsection}.

One way to judge the particle and field approaches is to compare the classical theories that are quantized to arrive at quantum field theories on the different approaches.  This comparison may help us to see which approach is built on the stronger foundation.

Before proceeding with that comparison, let me address a potential confusion: Our focus here is on classical theories that might yield quantum field theory upon quantization, not theories that arise in the classical limit as approximations to quantum field theory.  It is well-known that a classical theory of the Dirac field does not emerge in the classical limit as a macroscale approximation to quantum field theory (which \citealp[pg.\ 221]{duncan} explains as a consequence of Pauli exclusion).

Let us first compare our classical particle and field theories on their treatments of electron spin.  In a classical particle theory, we can endow the electron with an intrinsic angular momentum of magnitude $\frac{\hbar}{2}$ and an (always oppositely oriented) intrinsic magnetic moment of magnitude $\frac{e \hbar}{2 m c}$.\footnote{Here we are discussing a fully classical theory where the electron is modeled as a point particle that has an intrinsic ``spin'' angular momentum and an intrinsic ``spin'' magnetic moment.  In a Bohmian version of quantum mechanics or quantum field theory, you might include a point electron with these properties or without them (\citealp[ch.\ 9]{holland}; \citealp[ch.\ 10]{bohmhiley}).}  If we place the electron in an electromagnetic field, the existence of this intrinsic magnetic moment modifies the ordinary dynamics for a point charge.  The magnetic moment yields an additional force on the particle beyond the ordinary Lorentz force,
\begin{equation}
\vec{F}=q\vec{E}+\frac{q}{c}\vec{v}\times\vec{B}
\ ,
\label{particleforcelaw}
\end{equation}
that modifies its behavior in inhomogeneous magnetic fields.  The intrinsic magnetic moment also gives rise to a torque that alters the direction of the electron's angular momentum and---because we assume angular momentum and magnetic moment are always oppositely oriented---also alters the direction of the electron's magnetic moment.  To complicate things further, the electron's intrinsic magnetic moment acts as an additional source term in Maxwell's equations, producing a magnetic field around the electron.\footnote{\citet{barandes2019, barandes2021} develops a detailed classical relativistic theory of particles with intrinsic properties (like angular momenta and magnetic moments) interacting with the electromagnetic field.}

\begin{figure}[htb]
\center{\includegraphics[width=8 cm]{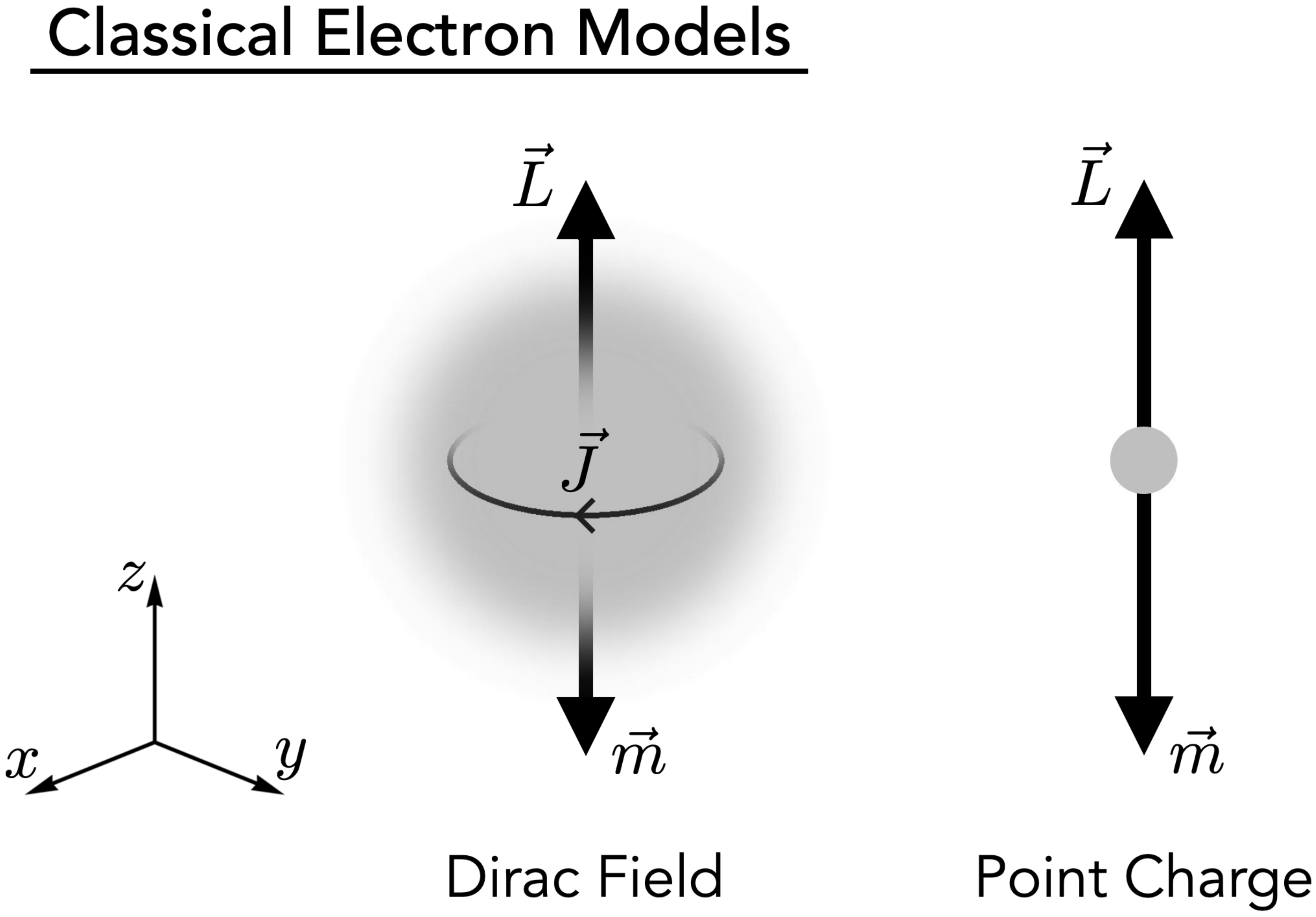}}
\caption{On the field approach, a z-spin up electron is modeled classically as a concentration of energy and charge in the classical Dirac field (where the charge density is shown here as a gray cloud).  The magnetic moment, $\vec{m}$, is generated by the current density, $\vec{J}$, describing the flow of negative charge in the opposite direction.  The angular momentum, $\vec{L}$, is generated by the momentum density (which is not shown here, but points opposite the current density).  On the particle approach, a z-spin up electron is modeled classically as a point charge with intrinsic angular momentum and magnetic moment.}
  \label{spinfigure}
\end{figure}

In classical Dirac field theory, the electron can be modeled as a cloud of charge with total charge $-e$ (composed entirely of positive-frequency modes).  The standard\footnote{These densities can be modified so that negative-frequency modes carry positive charge (as would be appropriate for representing positrons; see \citealp{positrons}), but we will not need to introduce that complication here as we are focused on electron spin.} charge and current densities for the Dirac field are
\begin{align}
\rho^q&=-e \psi^{\dagger}\psi
\label{diraccharge}
\\
\vec{J}&=-ec \psi^{\dagger}\vec{\alpha}\psi
\nonumber
\\
&=\frac{i e\hbar}{2 m}\left\{\psi^\dagger \beta \vec{\nabla} \psi - (\vec{\nabla} \psi^\dagger) \beta\psi\right\}\underline{ - \frac{e\hbar}{2 m} \vec{\nabla}\times(\psi^\dagger \beta \vec{\sigma}\psi)}+\frac{i e\hbar}{2 m c}\frac{\partial}{\partial t}(\psi^\dagger \beta\vec{\alpha} \psi)
\ ,
\label{diraccurrent}
\end{align}
which take the same form as the probability and probability current densities for the Dirac wave function in \eqref{diracdensities} (though in this context there are no such densities, as we are studying a classical field theory).  The second line of \eqref{diraccurrent} gives an expansion of the current density that can be arrived at using the free Dirac equation \eqref{thediracequation}.\footnote{This expansion of the current density is discussed in \citet{gordon1928}; \citet[pg.\ 321--322]{frenkel}; \citet[pg.\ 479]{huang1952}; \citet{ohanian}; \citet{howelectronsspin, smallelectronstates}.}  The underlined term in the expansion is the current associated with the electron's spin magnetic moment.  For a z-spin up Gaussian wave packet that is not too compact, the other terms are negligible and this term yields a current density describing a flow of charge around the z axis (figure \ref{spinfigure}).  This rotation of charge is responsible for the electron's magnetic moment.  Similarly, one can write out the electron's momentum density and identify a term associated with the electron's spin angular momentum.  In the kind of state just described, the momentum density would point opposite the current density, describing a flow of energy in the same direction as the flow of (negative) charge.  This rotation of energy is responsible for the electron's angular momentum.  In other states of the Dirac field, the flows of energy and charge may be more complicated, but we can always identify the contributions to these flows associated with spin magnetic moment and spin angular momentum.  In an external electromagnetic field, the force density $\vec{f}$ on the electron can be calculated from the standard Lorentz force law for continua,
\begin{equation}
\vec{f}=\rho^q\vec{E}+\frac{1}{c}\vec{J}\times\vec{B}
\ .
\label{forcelaw}
\end{equation}
The electromagnetic field sourced by the electron can be calculated from Maxwell's equations with the charge and current densities of the Dirac field acting as source terms.  For more on the picture of electron spin described in this paragraph, see \citet{ohanian, chuu2010, howelectronsspin, smallelectronstates, spinmeasurement}.

I see a number of advantages to the treatment of spin in this classical field model of the electron as compared to the classical particle model \citep[sec.\ 6]{spinmeasurement}.  First, in the field model one does not need to modify the Lorentz force law or add an additional torque law.  Second, there is no need to add intrinsic angular momenta or magnetic moments.  Angular momentum always and only results from the flow of energy (or you could say the flow of relativistic mass, which is proportional to energy).  Magnetic moments always and only result from the flow of charge.  This is a more unified account than the particle alternative where angular momentum sometimes arises from true rotation and sometimes is an intrinsic property of a point particle (with one type of angular momentum convertible into the other) and where magnetic fields can be produced by both moving charges and intrinsic magnetic moments.\footnote{The question as to whether it is only moving charges that produce magnetic fields has recently been discussed by \citet{fahy2022, griffiths2022}.}  Third, in the classical field model the electromagnetic field is simply sourced by charge and current densities---as in the usual way of writing Maxwell's equations.  By contrast, if we view the electron as a point particle with intrinsic magnetic moment we must modify (and complicate) Maxwell's equations to account for the role of this intrinsic magnetic moment as a source for the electromagnetic field.

Sometimes physicists say that the electron's angular momentum and magnetic moment cannot be generated by rotation because the electron is too small: if the electron's radius is much smaller than the Compton radius, $\frac{\hbar}{mc}$, there is no way to generate an angular momentum of $\frac{\hbar}{2}$ without the electron's mass rotating faster than the speed of light and no way to generate a magnetic moment of $\frac{e \hbar}{2 m c}$ without the electron's charge rotating faster than the speed of light.\footnote{Another common objection is that the electron's gyromagnetic ratio does not match the classical prediction.  But, that classical prediction assumes that mass and charge rotate at the same rate---which will not be the case for the mass and charge of the Dirac field \citep{howelectronsspin}.}  In brief, I take the solution to this puzzle to be that (in ordinary circumstances) the superposition of classical Dirac field states that forms the quantum state of the field is a superposition of states where the electron's relativistic mass (energy over $c^2$) and charge are not so tightly confined.  For example, in the hydrogen atom these might be states where the electron's relativistic mass and charge are spread throughout the atom's electron cloud---states where the electron is as big as the atom \citep[sec.\ 4.4]{quantumchemistry}.  It is possible to confine the electron's relativistic mass and charge so that they reside primarily within a sphere much smaller than the Compton radius, but it seems that when this is done the electron's relativistic mass becomes large and its magnetic moment becomes small, so that there is no need for either mass or charge to rotate superluminally (because there is enough relativistic mass to generate the ordinary angular momentum through rotation despite the small size of the mass distribution and because the rotation of charge does not have to yield the ordinary magnetic moment; \citealp{smallelectronstates}).  That being said, if we define the velocity of energy (or relativistic mass) flow as the energy flux density ($c^2$ times the momentum density) over the energy density, then it will exceed the speed of light in certain circumstances \citep{bb2021}.  More research is needed to better understand the flow of energy in such situations and whether an always slower-than-light velocity of energy flow can be found.\footnote{Although I hope that we can find a subluminal velocity of energy flow, I do not think this is a necessary condition for the picture of electron spin outlined here to be viable.  It may be better to focus on the densities of energy and momentum, recognizing that the above-defined velocity of energy flow is not always well-behaved.}  For charge flow, this problem does not arise.  If we define the velocity of charge flow as the current density divided by the charge density, it cannot exceed the speed of light for any state of the classical Dirac field.\footnote{See \citet[sec.\ 2b]{takabayasi1957}; \citet[sec.\ 10.4]{bohmhiley}; \citet[sec.\ 12.2]{holland}; \citet{howelectronsspin, smallelectronstates}.}

Let us now move on from spin and consider self-interaction.  On the field approach, the classical theory that will yield quantum electrodynamics upon field quantization is a theory where the Dirac field interacts with the electromagnetic field via Maxwell's equations (with the Dirac field as a source) and the Dirac equation (including interaction terms that were omitted in \eqref{thediracequation}).  On the particle approach, it is not clear what the full classical theory is supposed to be that will yield quantum electrodynamics upon quantization.  So far in this section, we have been thinking of it as a theory of classical point charges interacting with an electromagnetic field.  But, that is really a mixed theory of particles and fields which would presumably yield a mixed approach to quantum electrodynamics upon quantization, where a particle approach is taken for electrons and a field approach for photons.  This kind of approach was criticized in section \ref{PHOTONSsection}.  There are also a series of technical problems for this approach regarding self-interaction.  If you have a classical theory where point charges act as sources for the electric and magnetic fields via Maxwell's equations, then the electric field becomes infinitely strong as you approach a given charge.  First, this raises a problem of infinite self-energy because the integral of $\frac{E^2}{8\pi}$ in \eqref{EMenergy} diverges.  Second, this behavior of the electric field raises a problem for the Lorentz force law \eqref{particleforcelaw} dynamics because the electric field at the location of any point charge is ill-defined.  Third, we cannot simply remove self-interaction because we need particles to react to their own fields so that we can ensure conservation of energy and momentum when radiation is emitted---we need radiation reaction.  There are a plethora of strategies for addressing these problems.\footnote{For philosophical discussion of self-interaction in classical electrodynamics, see \citet{lange, frisch2005, earman2011}; \citet[sec.\ 4]{maudlin2018}; \citet{lazarovici2018, hartensteinhubert}.  For discussion in physics, see \citet{pearle, jackson1999, rohrlich2007, baez2021}.}  One can modify the Lorentz force law, change the way that point charges act as sources for the electromagnetic field, or remove the electromagnetic field and propose laws for direct interaction between particles across spatiotemporal gaps\footnote{There is much that could be said about the strengths and weaknesses of action-at-a-distance formulations of classical electrodynamics.  Briefly, note that such theories are non-local and violate both energy and momentum conservation (though there are ways of understanding locality and conservation that allow one to contest these apparent defects---see \citealp{lazarovici2018}).} (which could potentially yield a pure particle approach to quantum electrodynamics upon quantization).  The problems of self-interaction are difficult and there is no consensus as to how they should be resolved.  Some have hoped that resolving these problems in classical electrodynamics would help to remove problematic infinities that crop up in quantum electrodynamics (an idea that \citealp{feynman1965} discussed in his Nobel acceptance speech\footnote{See also \citet{blumjoas}; \citet[sec.\ 2.5.2 and 3.1]{blum2017}.}).

Let us now shift back to the field approach.  There remains work to be done to improve our understanding of electron self-interaction in a classical theory of interacting Dirac and electromagnetic fields and our understanding as to how the situation morphs upon field quantization.  That being said, there are good reasons to be optimistic.  At the classical level, the electron is an extended blob of charge.  The electric field does not become infinitely strong at any point and the total energy in the electromagnetic field is finite.  The Lorentz force law \eqref{forcelaw} always yields well-defined forces because the electromagnetic field never diverges.  Radiation reaction can potentially be explained by electromagnetic waves passing through the electron as it moves.\footnote{See \citet[sec.\ 2.2]{gravitationalfield}.}  Thus, the three problems raised in the last paragraph do not seem to be problems for the field approach.  However, there is another problem of self-interaction to consider: self-repulsion.

If a lone electron is modeled as a cloud of negative charge in the classical Dirac field, there should be a strong inwardly directed electric field throughout the electron that results in a strong outwardly directed force on each part of the electron.  In the face of this self-repulsion, what keeps the electron from exploding?  In a classical theory of interacting Dirac and electromagnetic fields there would be nothing to counteract this strong self-repulsion.  However, this classical tendency for an electron to explode appears to be eliminated in quantum electrodynamics, where Coulomb repulsion only occurs between distinct particles.

On the field approach, it is natural to wonder how electron self-repulsion is removed in the move from classical to quantum field theory.  Examining the classical theory of interacting Dirac and electromagnetic fields in the Coulomb gauge, one can isolate a Coulomb term in the Hamiltonian that includes the energy of self-repulsion for a single electron as well as repulsion between distinct electrons.  In the move to quantum electrodynamics, the Hamiltonian becomes an operator that directs the dynamics via a Schr\"{o}dinger equation like \eqref{superschrodinger}.  It is my understanding that every term in this Hamiltonian should be normal-ordered (with creation operators appearing to the left of annihilation operators).  It appears to be the normal-ordering of the Coulomb term in the Hamiltonian operator that deletes electron self-repulsion while keeping Coulomb interactions between distinct particles \citep{selfrepulsion}.

\subsection{Interactions and the Space of States}\label{STATESsection}

In the introduction, I mentioned the possibility of taking a deflationary approach where one sees the particle and field approaches as giving different ways of representing the same states.  On this approach, it would be wrong to say that fields are more fundamental than particles or that particles are more fundamental than fields.  Instead, one might say that there is a duality between particles and fields: quantum states can be written in either the particle basis or the field basis (as either particle wave functions or field wave functionals).  There will be problems for the particular case of photons, where a particle approach is not readily available (section \ref{PHOTONSsection}), but let us set those problems aside.

In support of the deflationary approach, one can find recipes for going from particle states to field wave functionals (\citealp[sec.\ 11.5]{bohmhiley}; \citealp[sec.\ 10.1]{hatfield}).  But, as I understand the situation, it is not always possible to go the other way.  The space of wave functionals appears to be larger than the space of particle wave functions, with the full set of possible particle wave functions being representable by a subset of the possible field wave functionals.  I have hedged my claims because the existing literature has not fully resolved these difficult technical issues.  Still, I see a potential advantage for the field approach here that I think is worth mentioning.  If the space of field states is indeed larger than the space of particle states, that would undermine the deflationary approach and would also cause problems for the particle approach (because the larger space of states seems to be needed when we consider quantum field theories that include interactions).

Haag's theorem is widely understood as teaching us that the (Fock space) wave functions over the various $n$-particle configuration spaces from section \ref{PARTICLEsection}, though they may be acceptable for free theories without interactions, cannot generally be used to describe states in quantum field theories that include interactions \citep{earman2006, fraser2008fate, ruetsche2012}.  The problem is that a ground state for the Hamiltonian of the interacting theory cannot be found within the original Fock space of particle states (describing superpositions of finitely many particles in different arrangements).  If we want a space of states that can be used for interacting quantum field theories, the particle wave functions from section \ref{PARTICLEsection} do not seem to be up to the task.

\citet[sec.\ 5]{baker2009} has argued that the above problem for particle wave functions also afflicts field wave functionals, because (he claims) the space of possible wave functionals is unitarily equivalent to the Fock space of possible particle wave functions.  However, the proof of this equivalence relies on the fact that the wave functionals discussed by \citet{baez1992}, \citet[sec.\ 6.3]{halvorson2007}, and \citet{baker2009} are restricted to range only over field configurations that are square-integrable, approaching zero at spatial infinity (the states of the classical field that look like normalizable single-particle wave functions).\footnote{Thank you to David Baker for clarifying this point in correspondence.} From the perspective of the field approach, there is no obvious physical reason to impose this restriction.  Classical field configurations do not need to be ``normalized.''  There are states of the classical Dirac and electromagnetic fields that go to zero as you approach spatial and infinity and others that do not.  I would think that a wave functional should assign amplitudes to all of these states.  That being said, imposing some kind of restriction like the restriction to square-integrable field configurations may be necessary in a mathematically rigorous development of the field approach.  \citet[sec.\ 4.5]{wallace2006} writes that the freedom for field configurations to ``have arbitrary large-distance boundary conditions ... interferes with the definition of the functional integral,'' a problem that can be overcome ``by imposing boundary conditions at infinity (such as a requirement that the [field configurations] be square-integrable).''\footnote{Difficulties related to functional integration were mentioned earlier in footnote \ref{measurefootnote}.}

If we do not impose the above restriction to square-integrable field configurations and instead allow our wave functionals to span over a wider range of classical field states than the authors above, then you would expect the space of wave functionals to be larger than the space of particle wave functions.  This is the conclusion that Jackiw reaches in his analysis of wave functionals.  After introducing wave functionals for bosons, \citet[pg.\ 4]{jackiw1987} writes: ``\dots our functional space is larger than any Fock space, indeed it contains all the inequivalent Fock spaces.  Put in another way, the Fock basis is not a complete basis for our functional space.''\footnote{See also \citet[pg.\ 88]{jackiw1990}.}  Given the concern raised by Wallace about functional integrals in the previous paragraph, one could challenge the viability of Jackiw's picture.  Still, I think the picture Jackiw paints is appealing and I would hope that the mathematical obstacles can be overcome.

There are a couple strategies that one might pursue to defend the particle approach from Haag's theorem and allow it to be applied to interacting quantum field theories.  One option is to render the number of classical degrees of freedom finite by imposing a high-momentum cutoff and working in a finite spatial region (moves you may already want to make for reasons relating to renormalization).\footnote{The reasons for introducing a high-momentum cutoff and a finite spatial region---and the costs that come with doing so---are discussed in \citet{wallace2006, wallace2021}; \citet[sec.\ 10.5]{duncan}; \citet{baker2016}; \citet[sec.\ 2]{deckert2020}.\label{cutofffootnote}}  Once this is done, the original particle Fock space can be used for both free and interacting theories (\citealp[sec.\ 10.5]{duncan}).  Another option is to retain an infinite number of classical degrees of freedom and to view the ground state of the interacting theory, at least for practical purposes, as a zero-particle state from which one can introduce single and multi-particle states as deviations---thus building a new space of particle wave functions for the interacting theory that is distinct from the space of particle wave functions for the free theory.\footnote{If the above kind of strategy works for introducing particle wave functions in interacting theories, there might be a way of combining the space of particle wave functions from the free theory with the various spaces used for different interacting theories to get a large space of states (that could perhaps be as big as the space of wave functionals).}  \citet{fraser2008fate} has argued against this kind of proposal.  \citet[pg.\ 211]{durr2020} have defended such a proposal in the context of the Dirac Sea, viewing interactions as inducing a change in sea level.

\section{Problems Facing the Field Approach}\label{CONSsection}

This section covers the problems arising for the field approach from the use of Grassmann numbers for fermionic fields, as well as the challenges involved in explaining how quantum particles emerge from quantum fields.  We have already touched on a number of other, arguably more minor, problems facing the field approach in the course of enumerating the advantages of the field approach over the particle approach: As was mentioned at the end of section \ref{PHOTONSsection}, there are open questions as to how positrons should be treated in a classical theory of interacting electromagnetic and Dirac fields so that you can most smoothly arrive at the standard theory of quantum electrodynamics upon field quantization.  In section \ref{CLASSICALsection}, we saw that there remains work to be done on understanding the flow of energy in the classical Dirac field (to complete the classical account of electron spin).  Section \ref{CLASSICALsection} ended by briefly presenting the absence of electron self-repulsion as a puzzle for the field approach and then summarizing a recently proposed solution.  Section \ref{STATESsection} mentioned the challenge of rigorously defining functional integration if wave functionals are allowed to span over a wide range of classical field configurations.

\subsection{Grassmann Numbers}\label{GRASSMANNsection}

The primary problems facing the field approach stem from the use of Grassmann numbers in wave functionals for fermionic fields, such as the Dirac field.\footnote{The problems with Grassmann numbers have led some to conclude that a field approach for fermions is either unavailable or unattractive.  See \citet[pg.\ 374]{bohm1987}; \citet[pg.\ 202]{durr2020}; \citet{struyve2010,struyve2011}; \citet[sec.\ 9.2]{wallace2021}.}  As was explained in section \ref{FIELDsection}, if we want the Dirac field operators to act on wavefunctionals as in \eqref{fieldoperator} and we want the field operators to anticommute \eqref{diracanticommutation}, then the classical Dirac field values must be anti-commuting Grassmann numbers---the classical Dirac field must be a Grassmann-valued field.  This leads to problems with classical interactions and problems with quantum probabilities.  Let us consider these in turn.

If the classical Dirac field is Grassmann-valued, then quantities like the field's charge density \eqref{diraccharge}, current density \eqref{diraccurrent}, energy density, and momentum density all fail to be real-valued or even complex-valued---they end up including Grassmann numbers and, in that sense, they are ``Grassmann-valued'' (\citealp[pg.\ 28]{bailinlove}; \citealp[appendix A]{positrons}).  When you consider interactions between the classical Dirac and electromagnetic fields (in a pre-quantization precursor to quantum electrodynamics), it is problematic that these quantities are not real-valued.  If the Dirac field's charge and current densities are not real-valued, how can they act as source terms in Maxwell's equations?  If the Dirac field's energy and momentum densities are not real-valued, how can energy and momentum be conserved in interactions where the electromagnetic field gains or loses energy or momentum?\footnote{Because it is ultimately the quantum field theory that needs to have a precise formulation, one might be willing to tolerate problems with energy and charge in the pre-quantization classical field theory so long as they do not deeply damage the post-quantization quantum field theory.  I would prefer, if possible, to start with a clear and consistent classical field theory.}

In addition to these problems for classical field theory before field quantization, there are problems for quantum field theory after field the quantization.  The use of Grassmann numbers for fermionic fields makes it difficult to interpret the wave functional's amplitude-squared as a probability density. \citeauthor{struyve2010} (\citeyear[sec.\ 9.2]{struyve2010}; \citeyear[sec.\ 3.3]{struyve2011}) raises concerns about defining a measure over the space of possible Grassmann-valued field configurations (as would be needed to integrate the probability density over subregions of field configuration space to get probabilities for certain kinds of field configurations).  Struyve also points out that the values of the wave functional itself are normally taken to include Grassmann numbers and thus not to be ordinary complex amplitudes.\footnote{See \citet{floreanini1988, jackiw1990}; \citet[sec.\ 10.3]{hatfield}; \citet{kiefer1994}.}  That is problematic because the wave functional's amplitude-squared would then not be real-valued (as a probability density must be).

These are difficult problems and it is not yet clear how to best navigate them.  In \citet[appendix A]{positrons}, I take a few steps down a particular path for avoiding the problems described above, though I am not certain it is the correct path.  That strategy begins by maintaining that our starting point for field quantization is a classical theory where you have the electromagnetic field interacting with a \emph{complex-valued} Dirac field $\psi^c_i(\vec{x})$ (thereby sidestepping the first set of problems regarding interactions in classical field theory).  As a mathematical tool, we can introduce a Grassmann-valued Dirac field $\psi^G_i(\vec{x})$, which is related to the complex-valued Dirac field by a one-to-one mapping such that specifying a configuration of the complex-valued Dirac field picks out a unique configuration of the Grassmann-valued Dirac field (and vice versa).  In quantum field theory, the Dirac field wave functional can then be viewed either as assigning amplitudes to configurations of the complex-valued or the Grassmann-valued Dirac field.  Setting the electromagnetic field aside, the wave functional can be written either as $\Psi[\psi^c,t]$ or $\Psi[\psi^G,t]$ (where the amplitude assigned by $\Psi[\psi^c,t]$ to a particular configuration of the complex-valued Dirac field at $t$ is the same as the amplitude assigned by $\Psi[\psi^G,t]$ to the corresponding configuration of the Grassmann-valued Dirac field at $t$).  If we take the field operator to multiply the wave-functional by the value of the Grassmann-valued Dirac field, $\widehat{\psi}_i(\vec{x})\Psi[\psi^c,t]=\psi^G_i(\vec{x})\Psi[\psi^c,t]$, then we can get the correct anticommutation relations while still viewing the wave functional as ranging over possible states for the complex-valued classical Dirac field (and thus the Grassmann-valued Dirac field turns out to be a useful mathematical tool to introduce).  Because the wave functional can be represented as ranging over configurations of the complex-valued classical Dirac field, we have a way of addressing Struyve's concern about defining a measure over the space of possible Grassmann-valued field configurations.\footnote{At least, we can define a measure here as easily as in the bosonic case.  That being said, there are challenges there (see footnote \ref{measurefootnote}).}  However, Struyve's challenge of finding an acceptable real-valued probability density remains.  The wave functional's amplitudes have not changed and thus the wave functional's amplitude-squared, $\Psi^{\dagger}\Psi$, still includes Grassmann numbers.  However, there may be a different quantity (derivable from the wave functional) that could serve as a real-valued probability density over the space of possible field configurations. \citet[pg.\ 245]{kiefer1994} point out that the inner product of a given wave functional with an eigenstate of definite field configuration is an ordinary complex number.  Thus, one could try taking the square of this quantity to be the probability density for that field configuration: $|\langle \psi^c | \Psi \rangle |^2$.

At this stage, more foundational work is needed to determine how Grassmann numbers can best be incorporated into (or excised from) a field wave functional approach to the nature of quantum states in quantum field theory.

\subsection{Getting Particles From Fields}

According to the field approach, quantum field theory is fundamentally a theory of fields.  That prompts the question as to why particle descriptions work as well as they do in the situations where they are successful.  How do quantum particles arise from quantum fields?

One way to address this general question is to analyze particular experimental situations with the goal of showing that the relevant wave functionals exhibit the appropriate particle-like behavior.  Proponents of wave functionals have studied the double-slit experiment, the Stern-Gerlach experiment, the photoelectric effect, and Compton scattering.\footnote{See \citet[pg.\ 363--373]{bohm1987}; \citet[ch.\ 11]{bohmhiley}; \citet[sec.\ 4]{kaloyerou1994}; \citeauthor{valentini1992} (\citeyear[sec.\ 4.1]{valentini1992}; \citeyear[pg.\ 54--55]{valentini1996}); \citet{spinmeasurement}.}  In the example of an idealized Stern-Gerlach experiment for the z-spin measurement of an x-spin up electron, the task would be to show that a wave functional for the electron that is initially centered on a classical Dirac field state describing the electron as spinning about the x-axis will evolve into a wave functional that is in a superposition of two separate pieces corresponding to the two possible measurement outcomes: one piece centered on a classical Dirac field state describing the electron as deflected upwards and spinning about the z-axis, and another piece centered on a classical Dirac field state describing the electron as deflected downwards and spinning about the z-axis in the opposite direction (see \citealp[sec.\ 7]{spinmeasurement}).  Further, one would need to show that the probability density over Dirac field configurations yields the correct probabilities for the two outcomes.

In addition to understanding particular experimental situations within the field approach, it is also important to understand why, in general, electrons can be described by four-component wave functions obeying the Dirac equation in relativistic quantum mechanics.\footnote{One might also wish to derive some quantum theory for the photon, but (as was discussed in section \ref{PHOTONSsection}) we have no theory like relativistic electron quantum mechanics for the photon---so the goalposts will look different for the photon.}  This theory has proved useful in many applications that go beyond non-relativistic quantum mechanics but do not require full quantum field theory, such as calculations of electron structure for atoms and molecules with heavy nuclei in quantum chemistry.\footnote{See \citet{desclaux2002}.}  How does the description of electrons in relativistic quantum mechanics approximate the behavior of wave functionals in quantum field theory?  If we can derive relativistic quantum mechanics from a field approach to quantum field theory, then it is straightforward to explain the success of non-relativistic quantum mechanics (as the derivation of the Pauli equation from the Dirac equation is well-known\footnote{See \citet[sec.\ 1.4]{bjorkendrell}; \citet[sec.\ 33]{lifshitzRQM}; \citet[sec.\ 10.4]{bohmhiley}; \citet[sec.\ 2.6]{ryder}; \citet{nowakowski1999}.}).

To derive relativistic quantum mechanics for a fixed number of electrons as an approximation to quantum field theory (on the field approach), we must be able to recover the states, the unitary dynamics, and the probabilities.\footnote{The details of this project will depend on one's preferred strategy for making the laws and ontology of quantum theories precise.  On the many-worlds interpretation, the task is as described above.  In an interpretation that includes some form of wave function collapse, one would have to propose a theory of wave functional collapse in quantum field theory and show that the collapse of the wave functional induces a satisfactory collapse of the particle wave function.  In a Bohmian field approach to quantum field theory where one supplements the wave functional with an actual field state evolving by a new equation of motion, one would have to show that the evolution of that field state leads to unique outcomes in quantum measurements.  One would not expect to (and would not need to) recover the point particles of elementary Bohmian quantum mechanics from the fields posited in the kind of Bohmian quantum field theory just described.}  In the absence of interactions, we can appeal to the mapping from particle wave functions to field wave functionals (mentioned in section \ref{STATESsection}) to see how the particle states of relativistic quantum mechanics can be reinterpreted as states of the Dirac field.  Continuing to set interactions aside, the Schr\"{o}dinger equation for the Dirac field wave functional should yield the free Dirac equation when applied to the aforementioned particle states.  To complete the story without interactions, one would need to show that the probability density over field configurations somehow yields the correct probability density over particle locations, at least when one considers hypothetical measurements.  I have not seen this issue addressed directly.

Including interactions with an external classical electromagnetic field or interactions between electrons (mediated by the electromagnetic field) will complicate the story.  In these contexts, it appears that there will not be an exact mapping from particle states to field states (section \ref{STATESsection})---though one can attempt to find field wave functionals that are fairly well-approximated by particular particle wave functions.  For the dynamics, the goal would be to start from the wave functional Schr\"{o}dinger equation of quantum electrodynamics (including interactions between the electromagnetic and Dirac fields) and then derive the appropriate version of the Dirac equation (including an external electromagnetic field or interactions between electrons) as an approximation to the field dynamics.  As in the free case, one would also need to get the probability density on the space of particle configurations from the wave functional's probability density on the space of field configurations.  I am not aware of much work on these problems that explicitly starts from field wave functionals, though there are related results one might adapt to this endeavor.

At this point, I see no reason to be pessimistic about the broad project of explaining particle-like behavior from a fundamental ontology of quantum fields (assuming the problems in section \ref{GRASSMANNsection} can be overcome).  That being said, there is important work to be done here that would help us to better understand the field approach to quantum field theory and its relation to quantum particle theories.

\section{Conclusion}\label{CONCLUSIONsection}

In this article, I have presented a snapshot of an ongoing debate between particle and field approaches to quantum field theory.  Being a snapshot, there is much left outside the frame and much that might change over time as research continues on these topics.  I have aimed for a snapshot that entices the viewer to explore further.  It would help to have more scholars engaging in foundational work on both approaches (and on alternatives).  The potential benefits of such work are significant.  First, to the extent that physics is aimed at understanding what exists (the ontology) and how the things that exist behave (the laws of nature), quantum field theory comes up short.  Clarity on the nature of quantum states would be significant progress towards precision about laws and ontology.  Second, settling whether we should take a particle or field approach to quantum field theory prepares the theory for the work that must be done to solve the quantum measurement problem.  This might be done via a many-worlds interpretation, a modification of the dynamics, or the addition of further ontology beyond the quantum state.  Adopting one of these strategies is necessary to make the ontology and laws of quantum field theory truly precise.  Third, there is a pedagogical payout to settling the question of particles versus fields.  Quantum field theory is notoriously difficult to teach and difficult to learn.  One problem with introductions to quantum field theory in current textbooks and courses is that they make the theory look unnecessarily alien, instead of being clear about how the theory relates to non-relativistic quantum mechanics and classical field theory.  A related problem is that these introductions are generally not explicit about what the laws of a particular quantum field theory are supposed to be and what kind of physical states are supposed to be governed by those laws.\footnote{See \citet{blum2017}.}  Explicitly and consistently taking either a particle or field approach would help the situation by allowing one to begin with a clear and intuitive description of quantum states and their dynamics that could be compared to the states and laws in other physical theories.  Progress that helps students learn and understand a theory can also help practitioners develop, apply, and extend the theory.  Fourth, seeking a precise formulation of quantum field theory may expose defects that can be remedied.  Einstein discovered special relativity by probing cracks in the foundations of classical electromagnetism.\footnote{See \citet[ch.\ 7]{lange}.}  Studying the foundations of quantum field theory could similarly lead to new ideas.

\vspace*{12 pt}
\noindent
\textbf{Acknowledgments}
Thank you to David Baker, Jacob Barandes, Jeffrey Barrett, Sean Carroll, Eddy Keming Chen, Maaneli Derakhshani, Benjamin Feintzeig, Mario Hubert, Dustin Lazarovici, Logan McCarty, Tushar Menon, David Mwakima, Ward Struyve, Roderich Tumulka, Jim Weatherall, and anonymous reviewers for helpful feedback and discussion.

\end{document}